# Signed Rank Chart For Tied Observations: An Application Of Deep Learning Models

Seyedeh Azadeh Fallah Mortezanejad, Ruochen Wang

*School of Automotive and Traffic Engineering, Jiangsu University, Zhenjiang, Jiangsu, China.*

∗ *Contact Ruochen Wang Email: wrc@ujs.edu.cn*

Abstract

Shewhart Control Charts (SCC)s are constructed under the assumption of normality and are widely recognized in statistical quality control by numerous researchers. Problems arise when the distribution of process data does not conform to a typical Normal Distribution (ND) or when there is insufficient evidence to confirm that the data has approximately ND. Additionally, in some processes, Tied Observations (TO)s are present. The resolution of the measurement device used to assess a quality characteristic can lead to rounding errors, as well as TOs. In many cases, SCCs prove inadequate. In this paper, we address the challenges of non-normal observations and rounding errors by developing a Shewhart Signed-Rank Control Chart (SS-RCC) based on the Wilcoxon statistic. We define a random variable for TOs and another for Untied Observations (UO)s. Subsequently, we approximate their distributions using a Scaled-Normal Distribution (SND) and apply a Deep Learning (DL) model to estimate the scale parameters of the SND for the Control Chart (CC). In practice, we calculate the Average Run Length ($ARL$) for specific cases using Johnson-type distribution benchmarks to illustrate the effects of ties and shifts in manufacturing processes.

# Introduction

One of the most renowned methods for detecting shifts in the manufacturing process is statistical quality control. Shifts in a process may result from changes in the mean or standard deviation. Unintentional shifts can lead to significant financial losses. Therefore, minimizing variations and diagnosing issues promptly and accurately are critical for preventing financial losses and saving time. In this context, Statistical Process Control (SPC) offers several important tools. The statistical CC was first introduced by Shewhart in 1924. However, SCCs are often unable to detect small shifts in processes. In such cases, a new design for a CC is necessary. To address this need, Cumulative Sum (CUSUM) CCs were first introduced by Page in 1954. Later, Exponentially Weighted Moving Average (EWMA) CCs were first introduced by Roberts in 1966. In recent years, modified versions of these charts have been utilized to accommodate new applications, as demonstrated in the works of Aslam et al. (2021), Xie et al. (2021), and Erem and Mahmood (2023) for CUSUM, as well as Xie et al. (2024) and Xue et al. (2024) for EWMA. Askari et al. (2023) focused on the optimal design of filters for CCs used in the monitoring process, addressing both time and frequency domain properties to enhance performance. By utilizing a Genetic Algorithm to minimize the ARL, the proposed optimal filter demonstrated superior speed in detecting shifts in process means. Jafarian-Namin et al. (2024) proposed an integrated model for statistical process monitoring, maintenance policy, and production, effectively addressing the challenges posed by autocorrelated data that can result in false alarms and delayed detection in traditional CCs. They applied an autoregressive moving average (ARMA) CC and implemented a particle swarm optimization algorithm to identify optimal decision variables, supported by industrial examples and a sensitivity analysis of the model parameters.

Two distinct types of changes occur in manufacturing processes: stochastic and non-stochastic changes. Stochastic changes are inherent and acceptable within processes, while non-stochastic changes arise from unexpected shifts. Upon detecting a point outside the control limits, technical specialists must investigate the source of the shift to minimize financial losses. Ghazanfari et al. (2008), Zarandi and Alaeddini (2010), and Amiri and Allahyari (2012) studied methods for accurately identifying change points in processes. If a CC is not selected appropriately, it may lead to false alarms or the failure to detect actual changes. Therefore, selecting a suitable CC is essential in this context. The primary assumption underlying the SCC is that the process data follows a ND. However, many processes do not conform to this distribution. Some processes may exhibit atypical patterns, such as trends, sudden shifts, systematic variations, cycles, or a combination of these characteristics, as discussed in Hachicha and Ghorbel (2012). In such cases, traditional SCCs are not appropriate. Consequently, several authors have addressed this issue and produced various papers, including works by Zhou et al. (2018), Anhøj and Wentzel-Larsen (2018), and Benková et al. (2023). Another significant aspect of SCCs is the estimation of their unknown parameters. Numerous articles, including those by Vargas (2003), Jones et al. (2004), Goedhart et al. (2017), and Jardim et al. (2020), focused on exploring methods for estimating the parameters SCCs.

In practical applications, processes or measurements often require rounding off results or are limited by the resolution of measurement devices, which can lead to a loss of true precision in recorded data. This situation introduces uncertainties and showy challenges for traditional calibration control methodologies. In this study, we propose a CC designed for TOs by leveraging the Wilcoxon signed-rank statistic, a widely used tool in non-parametric contexts.

The Wilcoxon signed-rank test, as introduced in Martín et al. (2016), is a non-parametric statistical test utilized to compare the medians of paired data or to assess differences between two related groups. This test is particularly valuable when the data do not satisfy the assumptions necessary for parametric tests, such as the paired t-test. Its advantages include robustness against outliers and non-normality in the data Joel et al. (2020); the absence of assumptions regarding a specific distribution, which makes it applicable to a diverse array of data types; suitability for ordinal data, where the precise values are less significant than the order or ranking of the observations Sayeedunnisa et al. (2018); and its effectiveness for paired data, where each observation in one group corresponds to a specific observation in the other group Li et al. (2021).

DL has numerous applications across various scientific fields, including computer science, computer vision, image and video processing, and even statistical quality control. For instance, Rezaei et al. (2024) presented an optimization algorithm for reducing prediction errors in time series using a DL network inspired by the visual cortex. The algorithm focused on improving the network structure, response speed, and neuron collaboration.

DL and neural networks with fewer hidden layers are commonly utilized in quality control literature due to their efficiency and accuracy. Shao and Hu (2020) addressed the challenge of recognizing systematic CC patterns in multiple-input multiple-output systems, which involved the complexity of simultaneous disturbances. It proposed the use of advanced machine learning classifiers, such as artificial neural networks and support vector machines, to effectively identify embedded mixture component concentration patterns. These classifiers had the potential to significantly enhance the recognition of complex CC patterns, ultimately improving process quality in intricate environments. Zaidi et al. (2023) enhanced pattern recognition for monitoring compositional data in SPC through the use of multivariate CCs. These charts often face challenges in Out-Of-Control (OOC) situations caused by shifts, outliers, or trends. They implemented a multi-layer feed-forward neural network. In this article, we employ a suitable DL model for our target data, specifically for a regression task. We use six fully connected hidden layers with a hyperbolic tangent activation function. For the regression task at the output layer, we apply a regression activation function.

In this paper, we are using Wilcoxon signed-rank statistic to build a suitable CC for TO data. Based on our current understanding, the challenge of monitoring TOs using a SS-RCC remains unresolved. In this study, we propose the application of the Wilcoxon signed-rank statistic and DL to address TOs. When there are no ties present, the ND approximates the distribution of this statistic quite well. However, when ties do appear in the data, we first remove their presence from the observations using a specific technique, and then recalculate the necessary parameters for this situation. The distribution of these data resembles a ND, but we reconstruct the height, location, and width of the ND using three new parameters. Based on these operations, we define a SND. In this case, the occurrence probability of each point in the sample space is calculated discretely, ensuring that the total of these probabilities equals one, thereby maintaining the property of density functions. Another challenge that remains is how to accurately estimate these three new parameters. To address this, we develop a suitable DL model. Finally, for various scenarios created by Johnson-type distributions, we first calculate the required parameters and then establish appropriate control limits. To assess the effectiveness of these control limits, we use *ARL* as a measure of performance.

In the simulation section of this paper, we utilize a Johnson-type distribution. Johnson-type distribution, as introduced in Johnson (1949), is a transformation of the ND function that depends on four parameters: $\zeta$, $\vartheta > 0$, $\iota$, and $\xi > 0$. This distribution encompasses various families of distributions, whose domains can be either bounded or unbounded. The cumulative distribution function (c.d.f.) for an unbounded random variable $Y$ at a point $y$ is defined as:

$$F_{S_U}(y) = \Phi\left(\zeta + \vartheta \sinh^{-1}\left(\frac{y-\iota}{\xi}\right)\right), \quad y \in (-\infty, \infty), \tag{1.1}$$

where $\Phi(\cdot)$ represents the c.d.f. of the standard ND. When the random variable is bounded, the definition is different:

$$F_{S_B}(y) = \Phi\left(\zeta + \vartheta \ln\left(\frac{y-\iota}{\iota+\xi-y}\right)\right), \quad y \in [\iota, \iota+\xi]. \tag{1.2}$$

The contributions of this paper are outlined as follows: in Section 2, we develop a SS-RCC specifically tailored for datasets with unknown distributions and free from rounding errors; in Section 3, we address TOs arising from the resolution limitations of measurement devices in a process and introduce SNDs for the unknown distribution of TOs; in Section 4, we calculate the parameters of the SNDs for the simulation study using a DL model, and compute the $ARL$s for various scenarios, both with and without ties, to illustrate the impact of tie presence on detecting shifts in manufacturing processes; in Section 5, we present the conclusions and insights derived from this study.

# Wilcoxon Signed-Rank Chart For UO

There are situations in which observations do not contain any ties. For example, this can occur when the measuring instrument is highly precise and does not produce rounding errors. However, even with this level of accuracy, the distribution may remain unspecified. In these scenarios, it is not suitable to apply the standard SCC. Instead, one can use the straightforward SS-RCC for regular datasets. Consequently, our primary emphasis is on identifying the correct chart in the absence of ties.

Let's consider $X$ as the quality characteristic at time $t$, following an unknown continuous distribution denoted by $F_X(.\,|v)$. Here, $v$ represents a location-scale parameter that we aim to monitor. The process is deemed In-Control (IC) when $v = v_0$, but it transits to an OOC state when $v = v_1$. The corresponding plots can be found in Figure 1. Without loss of generality, let us assume that $X$ is an unbiased estimator of the parameter $v$. The observations of the quality characteristic $X$ are denoted by $\{X_{t,k}\}_{k=1,\ldots,n}$. To apply the Wilcoxon signed-rank statistic, we consider a scenario where at time $t = 1, 2, \ldots$ we have $n$ discrete independent random variables $\{S_{t,k}\}_{k=1,\ldots,n}$ defined on set $\{-1, +1\}$ with corresponding probability mass functions $P(S_{t,k} = -1) = 1 - p$ and $P(S_{t,k} = +1) = p$, such that $S_{t,k} = sign(X_{t,k} - v_0)$ for $k = 1, \ldots, n$. Let $SR_t$ be the random variable defined as:

$$SR_t = \sum_{k=1}^{n} k\, S_{t,k}, \tag{2.1}$$

where its support set is $\mathbb{S} = \{-\frac{n(n+1)}{2}, -\frac{n(n+1)}{2}+2, \ldots, 0, \ldots, \frac{n(n+1)}{2}-2, \frac{n(n+1)}{2}\}$. It is worth mentioning that the length of steps in $\mathbb{S}$ is 2. The reason is as follows: be explained with an example where all $S_{t,k}$ values are positive except for $k=1$. Therefore, $SR_t = \left(\frac{n(n+1)}{2} - 1\right) - 1$. The change in the statistic $SR_t$ when adding or removing a particular $k$ due to a change in the sign (from positive to negative or vice versa) results in a shift of $\pm 2k$ in the value of $SR_t$.

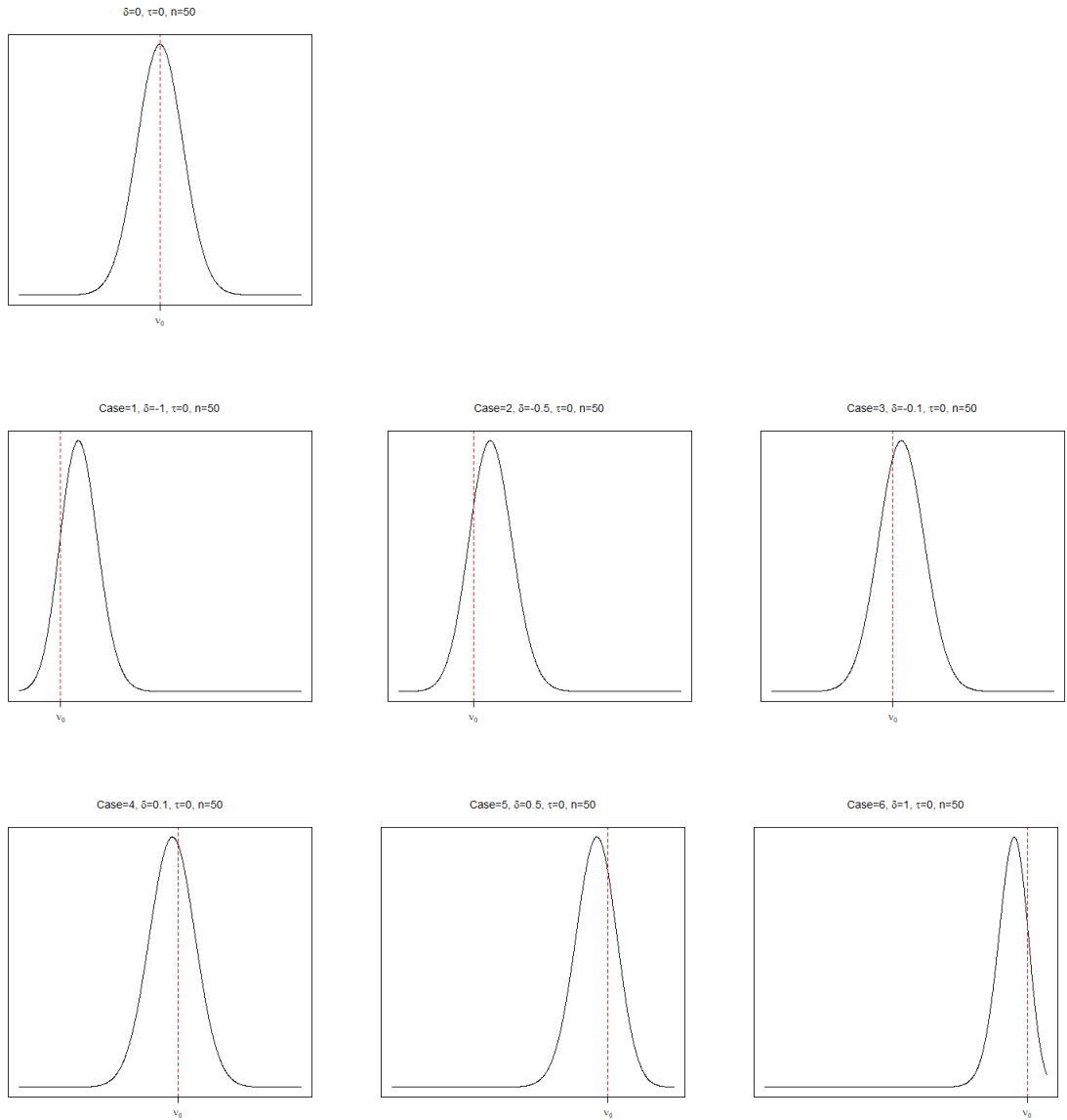

Figure 1: The top plot illustrates the IC scenario for UOs with zero values of $\tau$ and $\delta$. The subsequent curves represent OOC situations with various shift values indicated above each plot. These OOC curves are derived from a range of benchmark Johnson-type distributions defined in Table 1.

We denote the expectations and variance of $S_{t,k}$ as:

$$\begin{cases} m_1(S_{t,k}) = E(S_{t,k}) = 2p - 1 \\ m_2(S_{t,k}) = E(S_{t,k}^2) = 1 \\ \mu_2(S_{t,k}) = E\left[\left(S_{t,k} - m_1(S_{t,k})\right)^2\right] = 4p(1-p) \end{cases},$$

where $E(\cdot)$ means the mathematical expectation. Based on these values, we can derive:

$$m_1 = m_1(SR_t) = E(SR_t) = m_1(S_{t,k}) \underbrace{\sum_{k=1}^n k}_{\frac{n(n+1)}{2}} = \frac{n(n+1)(2p-1)}{2}. \tag{2.2}$$

We nominate $\mu_2 = \mu_2(SR_t)$, and then:

$$\begin{aligned} \mu_2 &= E(SR_t^2) - m_1^2 \\ &= \frac{n(n+1)(2n+1)}{6} + (2p-1)^2 \left(\frac{n^2(n+1)^2}{4} - \frac{n(n+1)(2n+1)}{6}\right) - m_1^2 \\ &= \frac{2np(1-p)(n+1)(2n+1)}{3}. \end{aligned} \tag{2.3}$$

From Equations (2.2) and (2.3), we compute the probability mass function (p.m.f.) $f_{SR_t}(s|n,p)$ of $SR_t$ using its c.d.f. $F_{SR_t}(s|n,p)$:

$$f_{SR_t}(s|n,p) = F_{SR_t}(s|n,p) - F_{SR_t}(s-2|n,p), \tag{2.4}$$

where $F_{SR_t}(s|n,p)$ for $s \in \mathbb{S}$ is estimated by the standard ND c.d.f. $\Phi(\cdot)$:

$$F_{SR_t}(s|n,p) \simeq \Phi\left(\frac{s+0.5-m_1}{\sqrt{\mu_2}}\right). \tag{2.5}$$

The corresponding control limits $(UCL, LCL)$ are $(-C, C)$ with the center line $CL = 0$, where $C \in \{2, \ldots, \frac{n(n+1)}{2}\}$ or $C \in \{1, \ldots, \frac{n(n+1)}{2}\}$ depends on $\frac{n(n+1)}{2}$ for even or odd numbers, respectively. The value of $C$ depends on the technician's choice of confidence level $100(1-\alpha)\%$, which is usually chosen as $99.73\%$, where $\alpha$ is the first type of error. A process is declared to be IC if $-C < SR_t < C$, and otherwise, it is considered OOC. The first type of error, $\alpha$, for UO is as follows:

$$\tag{2.6}$$

$$1 - \alpha = P(-C < SR_t \leq C - 2|v = v_0) \simeq \Phi\left(\frac{C-1.5-m_1}{\sqrt{\mu_2}} \Big| v = v_0\right) - \Phi\left(\frac{-C+0.5-m_1}{\sqrt{\mu_2}} \Big| v = v_0\right).$$

The value that satisfies this equation is the desired $C$. However, the problem is that we cannot solve for $C$ analytically. It is costly and can only be solved numerically. The second type of error $\beta$ for UO is:

$$\begin{aligned} \beta &= P(-C < SR_t \leq C - 2|v = v_1) \\ &\simeq \Phi\left(\frac{C-1.5-m_1}{\sqrt{\mu_2}} \Big| v = v_1\right) - \Phi\left(\frac{-C+0.5-m_1}{\sqrt{\mu_2}} \Big| v = v_1\right). \end{aligned}$$

Pay attention to the value of $p$ used to calculate $m_1$ and $\mu_2$. In the following, we see that $p$ is equal to 0.5 for the IC process when $\nu = \nu_0$, and $p \neq 0.5$ when $\nu = \nu_1$. Therefore, we can compute $C$ with the desired $\alpha$ using $p = 0.5$ for the IC process.

The run-length is defined as the number of samples taken before we either conclude that the process is OOC or incorrectly accept that it is IC when it is actually OOC. The concept of run-length distribution refers to the number of samples collected before a specific event occurs. The run-length distribution in SS-RCCs follows a geometric distribution with parameter $\alpha$ or $1 - \beta$. IC $ARL$ notated as $ARL_0 = \frac{1}{\alpha}$ refers to the average number of samples taken before a signal indicates an OOC condition when the process is actually in control. OOC $ARL$ computed as $ARL_1 = \frac{1}{1-\beta}$ refers to the average number of samples taken before a signal indicates that the process is OOC when it actually is. If $\alpha = 0.0027$ the minimum acceptable $ARL_0$ is 370.370. The larger the $ARL_0$, the more robust the CC becomes. The lower the $ARL_1$, the more accurate the alarm generated by the CC.

It is important to note that the formulas for $\alpha$ and $\beta$ are accurate for UOs, indicating that the possible set of $S_{t,k}$ is $\{-1, +1\}$, as mentioned earlier. However, in the presence of ties, the set becomes $\{-1, 0, +1\}$, necessitating new formulas for $\alpha$ and $\beta$. In the following section, we focus on TOs and determine the appropriate formulas for the errors.

# Wilcoxon Signed-Rank Chart For TO

The procedures outlined in the previous section are applicable when there are no ties present. However, if ties exist, this approach is not appropriate. In the presence of ties, 0 becomes one of the possible events for $S_{t,k}$, necessitating a different approach. Therefore, it is essential to first define ties. Ties often occur due to the resolution of measurement devices, where the true values of the desired quality characteristic $X_{t,k}$ of a process are not directly recorded. Instead, a measured value $X'_{t,k}$ is observed, which may not be equal to $X_{t,k}$. In such cases, various types of errors can occur, and the comprehensive model for these errors is:

$$X'_{t,k} = \lfloor \frac{1}{\eta}(\alpha_1 + \alpha_2 X_{t,k} + \epsilon_{t,k}) + 0.5 \rfloor \eta.$$

The parameter $\eta$ represents the quality parameter of device resolution, while $\alpha_1$ and $\alpha_2$ denote bias linearities. The term $\epsilon_{t,k}$ signifies precision error, and $\lfloor \cdot \rfloor$ denotes the floor function. The presence of $\eta$ leads to rounding-off errors in measurements. The minimum shift in the quality characteristic that a measurement device can accurately indicate is determined by the resolution of the measurement device. When a point is recorded as $X'_{t,k} = x$, it does not represent the exact quality value; instead, the precise quantity $X_{t,k}$ falls within the interval of $(x - \frac{\eta}{2}, x + \frac{\eta}{2}]$.

For instance, in a manufacturing process, the length of a product may be the desired quality characteristic, with a target value of $\nu_0 = 15\ cm$ (centimeter) and a measurement device resolution of $0.1\ cm$. In this scenario, the set of possible observations is denoted as

{...,14.8,14.9,15,15.1,15.2,...}. If the actual value of a product is $X_{t,k} = 15.04\ cm$, the observed quantity $X'_{t,k}$ may be recorded as 15. Consequently, ties are introduced, rendering the variable $S_{t,k}$ defined in the previous section inadequate due to the inclusion of 0 as one of the conceivable events. Therefore, a new discrete random variable, $\{S'_{t,k}\}_{k=1,...,n}$, must be defined, with a feasible set of $\{-1,\ 0,\ +1\}$ and a corresponding probability distribution:

$$P(S'_{t,k} = -1) = p_{-1},\ P(S'_{t,k} = +1) = p_{+1}, \tag{3.1}$$

and

$$P(S'_{t,k} = 0) = 1 - p_{-1} - p_{+1} = p_0. \tag{3.2}$$

In this context, it is advisable to eliminate the zeros that lead to ties. Therefore, a new statistic $SR_{t,N} = \sum_{k=1}^{N} k\, S'_{t,k}$ needs to be defined, where $N$ is a random variable following a binomial distribution with the parameters $n$ and $1 - p_0$. In present of ties, the possible set of $SR_{t,N}$ is $s \in \{-\frac{n(n+1)}{2}, -\frac{n(n+1)}{2} + 1, ..., 0, ..., \frac{n(n+1)}{2} - 1, \frac{n(n+1)}{2}\}$. However, when ties are removed, the steps change to 2, similar to the scenario when there were no ties present. Consequently, the current definition of the Wilcoxon rank statistic excludes all ties, requires the establishment of a new probability distribution for $S'_{t,k}$:

$$\pi_{-1} = P(S'_{t,k} = -1) = \frac{p_{-1}}{p_{-1} + p_{+1}},$$

$$\pi_{+1} = P(S'_{t,k} = +1) = \frac{p_{+1}}{p_{-1} + p_{+1}}.$$

This definition applies in scenarios where ties are present. We now aim to reevaluate the necessary parameters of TOs using $SR_{t,N}$, where $N$ is already a random variable. To achieve this objective, we begin by calculating specific expectations of $N$ through the probability generating function of the binomial distribution, as detailed in Appendix A. Subsequently, our results depend on these expectations.

Below, we identify several key parameters of $S'_{t,k}$ that are essential for calculating the skewness and kurtosis of $SR_{t,N}$:

$$\begin{cases} m_1(S'_{t,k}) = m_3(S'_{t,k}) = \pi_{+1} - \pi_{-1} \\ m_2(S'_{t,k}) = m_4(S'_{t,k}) = 1 \\ \mu_2(S'_{t,k}) = 4\pi_{+1}\pi_{-1} \\ \mu_3(S'_{t,k}) = E\left[\left(S'_{t,k} - m_1(S'_{t,k})\right)^3\right] = 8\pi_{+1}\pi_{-1}(\pi_{-1} - \pi_{+1}) \end{cases}.$$

We now have the necessary parameters for $SR_{t,N}$. Therefore, we calculate $m'_1$, $m'_2$, and $m'_3$ using dual conditional expectation:

$$m'_1 = m'_1(SR_{t,N}) = E\left(E\left[\sum_{k=1}^{N} k\, S'_{t,k}|N\right]\right) = \frac{\pi_{+1} - \pi_{-1}}{2} E(N^2 + N),$$

$$m'_2 = EE\left(\frac{1}{3}N^3 + \frac{1}{2}N^2 + \frac{1}{6}N\right) + (\pi_{+1} - \pi_{-1})^2 E\left(\frac{1}{4}N^4 + \frac{1}{6}N^3 - \frac{1}{4}N^2 - \frac{1}{6}N\right),$$

$$m'_3 = \frac{\pi_{+1} - \pi_{-1}}{4} E(2N^5 + 3N^4 - N^2) + \frac{(\pi_{+1} - \pi_{-1})^3}{8} E(N^6 - N^5 - 3N^4 + N^3 + 2N^2).$$

Detailed information is provided in Appendix A, along with some closed-form summations in Appendix B. Additionally, $\mu'_2$, $\mu'_3$, and $\mu'_4$ are calculated in Appendix A. Then, we calculate the Fisher's and Pearson's coefficients of skewness and kurtosis for $SR_{t,N}$, denoted as $\gamma'_1$ and $\gamma'_2$, respectively:

$$\gamma'_1 = \frac{\mu'_3}{\mu'^{3/2}_2}, \quad \gamma'_2 = \frac{\mu'_4}{\mu'^{2}_2}.$$

Skewness measures the degree of asymmetry in a distribution, while kurtosis assesses the degree of peakeness. If $\gamma'_1 > 0$, the distribution is skewed to the right; conversely, if $\gamma'_1 < 0$, the distribution is skewed to the left. When $\gamma'_2 > 3$, the distribution exhibits a high peak and is referred to as leptokurtic. In contrast, if $\gamma'_2 < 3$, the distribution has a low peak and is described as platykurtic. When $\gamma'_2 = 3$, the distribution is termed mesokurtic. The ND has $\gamma'_1 = 0$ and $\gamma'_2 = 3$, Jambu (1991). Another kurtosis measure is Fisher's coefficient, which is represented as $\beta_2 = \gamma'_2 - 3$. In a ND, $\beta_2$ equals 0.

The next stage is to make an approximation for $SR_{t,N}$ distribution. We propose a scaled version of ND with mean $m'_1$ and variance $\mu'_2$, when $\pi_{+1} \neq 0.5$:

$$f_{SR_{t,N}}(s) \simeq \int_{s-0.25}^{s+0.25} a\,\gamma'_2 \phi\left(\frac{x - b\gamma'_2}{c\gamma'_1}\Big|m'_1, \mu'_2\right) dx, \quad s \in \mathbb{S},\ c, \gamma'_1 \neq 0,$$

where $\phi(\cdot)$ is the probability density function (p.d.f.) of ND. The corresponding c.d.f. is calculated by:

$$F_{SR_{t,N}}(s) \simeq \sum_{\substack{i=-\frac{n(n+1)}{2} \\ i \in \mathbb{S}}}^{s} f_{SR_{t,N}}(i), \quad s \in \mathbb{S}.$$

$a$, $b$, and $c$ adjust the height, location, and width of the ND, with a mean of $m'_1$ and a variance of $\mu'_2$. We refer to the newly adjusted distribution as SND. The larger the value of $a$, the higher the distribution, and is always positive. As $b$ increases, the location of SND shifts to the right; remember, when there are negative shifts, $b < 0$. The larger the absolute value of $c$, the wider the SND becomes, and it is negative for larger negative shifts. These three recently defined parameters change in different scenarios and are difficult to evaluate. Therefore, we propose a DL model for their computation. The information we have indicates that these parameters depend on the values of $\pi_{+1}$, $m'_1$, $m'_2$, $m'_3$, $m'_4$, $\mu'_2$, $\mu'_3$, $\mu'_4$, $\gamma'_1$, and $\gamma'_2$. Additionally, they are influenced by the increments of $n$. In the following section, we develop a DL model for $n = 20$ and $n = 50$ in shifted settings.

When $\pi_{+1} = 0.5$, we use (2.4) and (2.5) with corresponding $m'_1$ and $\mu'_2$ for TO. It is important to note that when $\pi_{+1} \neq 0.5$, both $c$ and $\gamma'_1$ are non-zero values. For the calculation of the related control limits $(-C, C)$, we apply (2.6) with the substitution of $m'_1$ and $\mu'_2$ because, in the absence of shifts, $\pi_{+1} = 0.5$. In the following section, we demonstrate this concept for various scenarios. In the presence of unwanted shifts, $\pi_{+1}$ becomes unequal to 0.5. The first and second types of errors for TOs are as follows:

$$\begin{aligned} \alpha &= 1 - F_{SR_{t,N}}(C|v=v_0) + F_{SR_{t,N}}(-C|v=v_0), \\ \beta &= F_{SR_{t,N}}(C-2|v=v_1) - F_{SR_{t,N}}(-C|v=v_1). \end{aligned}$$

Without loss of generality, suppose that for every shift in processes, we have $v_1 = v_0 + \delta\sigma$, where $\delta$ represents the standardized shifts and $\sigma$ is the standard deviation of $X_{t,k}$. Then, the probabilities given in equations (3.1) and (3.2) for the occurrence of the components in the set $\{-1, 0, +1\}$ in the presence of ties and shifts are:

$$\begin{cases} p_{-1} &= F\left(-\frac{\tau}{2} - \delta\right) \\ p_0 &= F\left(\frac{\tau}{2} - \delta\right) - F\left(-\frac{\tau}{2} - \delta\right) \\ p_{+1} &= 1 - F\left(\frac{\tau}{2} - \delta\right) \end{cases} \quad (3.3)$$

where $\tau = \frac{\eta}{\sigma}$ is the standardized resolution, which indicates ties in processes, the amount of $\delta$ represents shifts. The related plots are shown in Figure 2 for a better understanding of the impacts of ties and shifts on the distribution of $SR_{t,N}$. The function $F(\cdot)$ is the distribution function defined in equations (1.1) and (1.2). When $\delta = 0$ is substituted into the recent equations, the IC values of $p_{-1}$, $p_0$, and $p_{+1}$ are obtained. Similarly, when $\tau = 0$ is applied in the formulas, the corresponding probabilities of UOs are determined. Thus, the purpose of these formulas is to calculate the probabilities of $p_{-1}$, $p_0$, and $p_{+1}$ in the presence of ties, shifts, or a combination of both. This enables us to determine $\alpha$ and $\beta$, and subsequently, the $ARL_0$ and $ARL_1$. We provide examples of these functions in Table 1 to clarify their application for simulation purposes in the next section. Then, $p_{-1}$, $p_0$, and $p_{+1}$ are derived from six different symmetric Johnson-type distributions based on the known values of $\tau$ and $\delta$.

## Detecting Shift In Practice

In previous sections, we illustrate SS-RCCs for UOs and TOs. Furthermore, we presented a method for modelling the ties and shifts in manufacturing processes. In this section, we calculate $ARLs$ in practice. To do this, we require the primary probabilities $p_{-1}$, $p_0$, and $p_{+1}$ calculated for six different symmetric Johnson-type distribution benchmarks listed in Table 1. These calculations are conducted under various conditions involving UOs and TOs, as shown in

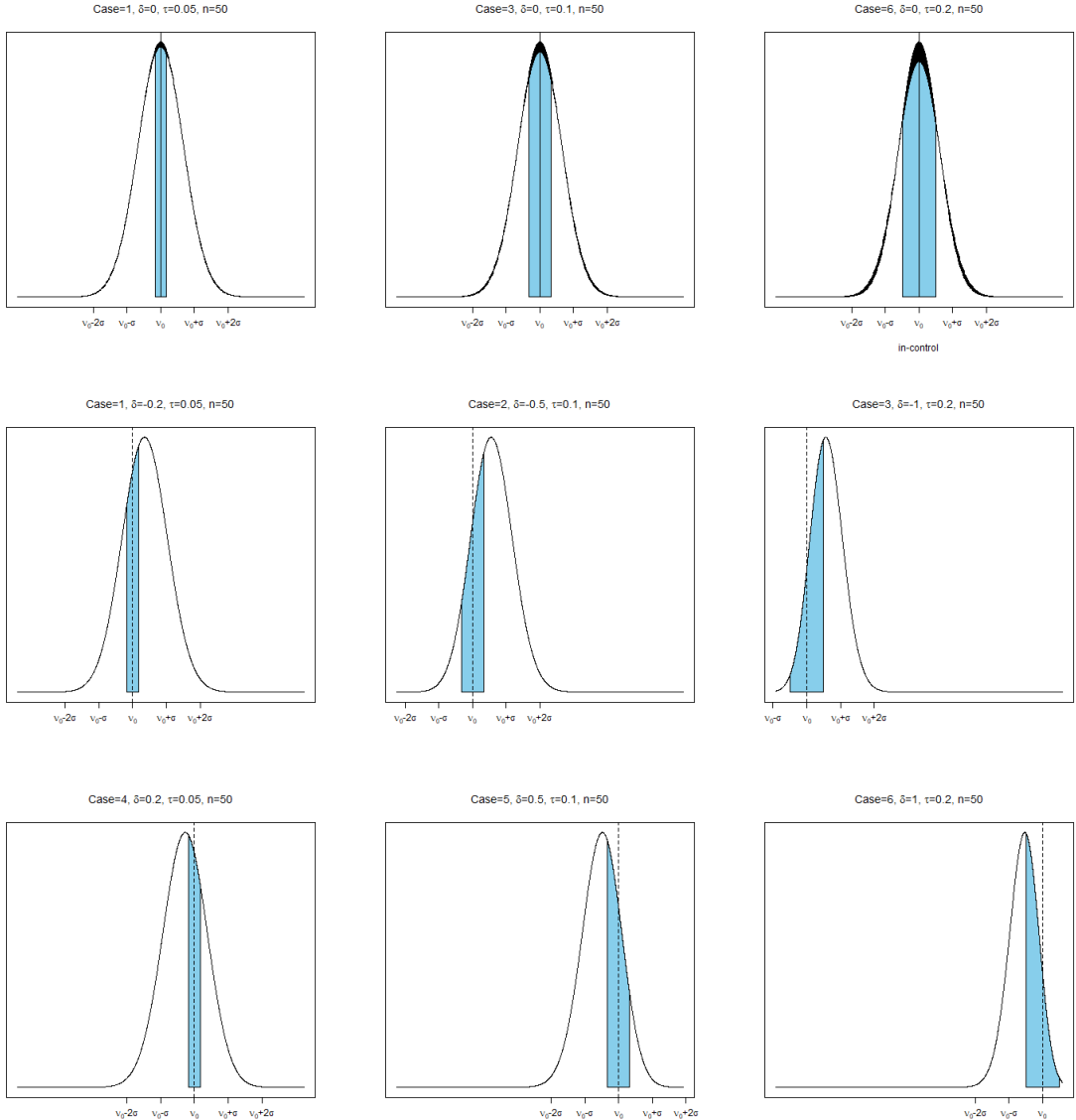

*Figure 2: The first line of plots illustrates IC scenarios with ties, benchmarked defined by the Johnson-type distributions presented in Table 1 for cases 1, 3, and 6. Additional curves represent OOC situations with varying shifts. The coloured areas indicate the probabilities of encountering ties within the processes.*

Table 2, enabling us to compare the effects of measurement resolution. The advantage of using this method to generate probabilities is that all probabilities are calculated under specific conditions, and none of them are equal, except in cases where there are no ties and shifts.

In Table 2, nearly all values of $p_{+1}$ and $p_{-1}$ of the IC situation with $\delta = 0$ are equal with different $\tau$s and cases. In the case of TOs, $\pi_{+1}$ and $\pi_{-1}$ both equal 0.5, which corresponds to the IC UO settings. For the calculations of $ARL_0$, we utilize an ND with a mean of $m'_1$ and a variance of $\mu'_2$, derived from the TO formulas. The first six rows present the probabilities of UOs; therefore, we apply the ND with a mean and variance of $m_1$ and $\mu_2$ for the calculations of both $ARL_0$ and $ARL_1$. The first two rows in Figure 7 display density plots of UOs for various scenarios, as indicated for each plot. They consist of two curves: one representing the simulation densities and the other representing the ND densities. The simulated densities are based on 1,000,000 repetitions. As a result of the comparison, in the absence of ties, the density can be approximated using an ordinary ND with different parameters for either IC or OOC situations. The situation becomes much more challenging in cases involving TOs for OOC scenarios. We discuss the steps to approximate the densities in the following.

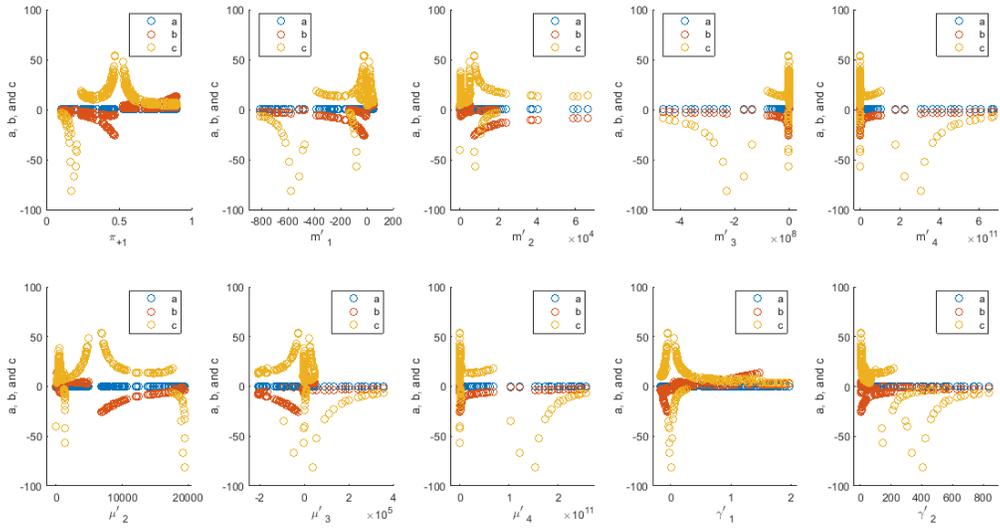

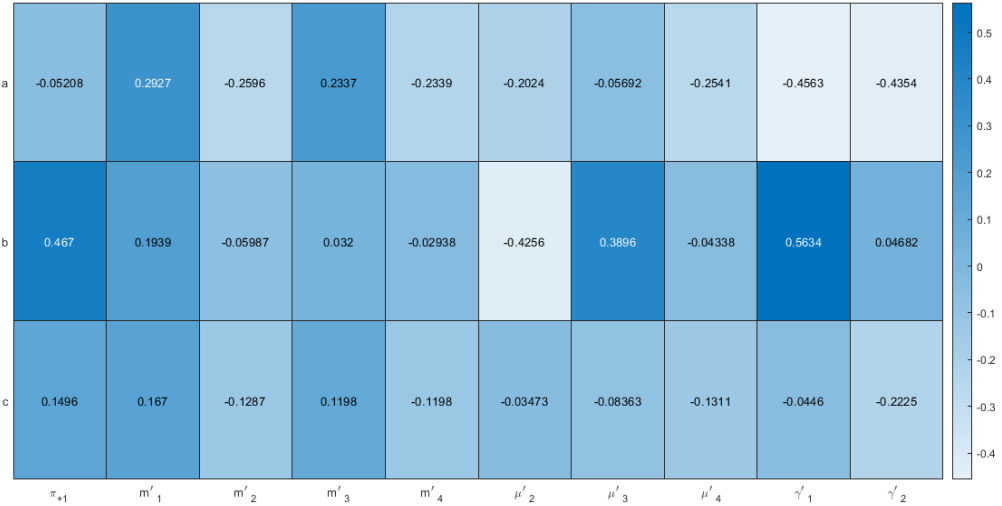

*Figure 3: At the top, there are scatter plots of a, b, and c with respect to the parameters indicated on the x-axes of each plot, allowing us to observe the available dependencies. Below, there is a correlation map of those parameters.*

Using the probabilities in Table 2 for OOC and TOs, we need to calculate all the required parameters, such as $m'_1$, $m'_2$, etc. Here, we specifically considered cases where the sample size is either $n = 20$ or $n = 50$. After calculating the necessary parameters, we manually determined suitable values for $a$, $b$, and $c$. As a result, we created a dataset consisting of 288 samples with 11 input variables and 3 different output variables.

To ensure the existence of dependencies among the parameters, we plot a scatter diagram of $a$, $b$, and $c$ in Figure 3. It is evident that changes in parameters such as $m'_1$, $m'_2$, and others lead to variations in the values of $a$, $b$, and $c$. In other words, the values of $a$, $b$, and $c$ are dependent on the levels of those parameters. Below Figure 3, we exhibit the correlation map, highlighting their varying degrees of interdependence.

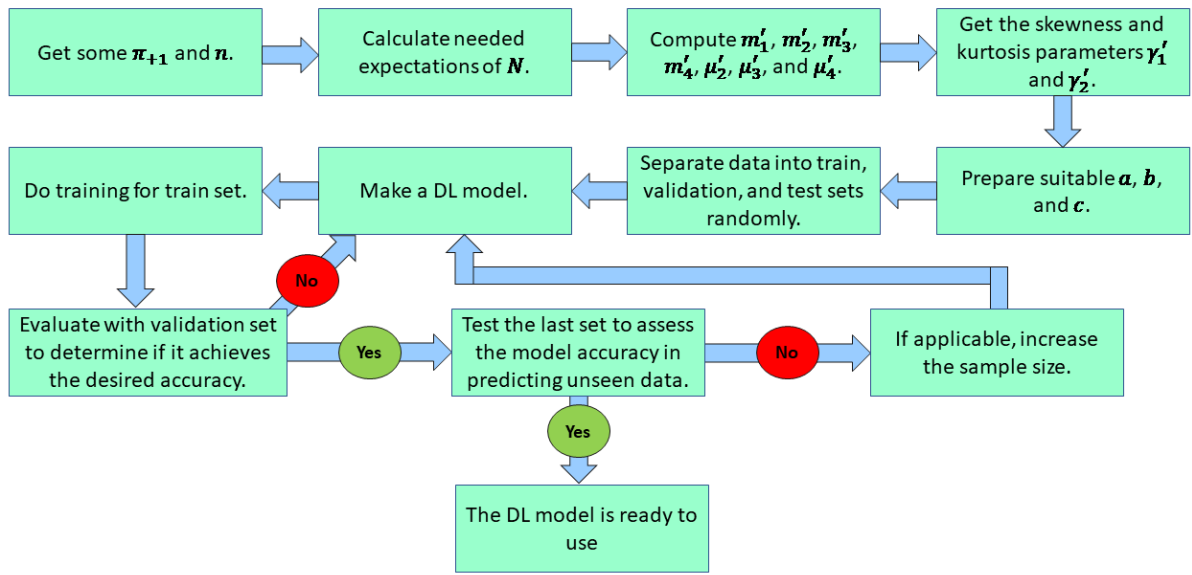

Figure 4: The algorithm of handling TOs in a DL model.

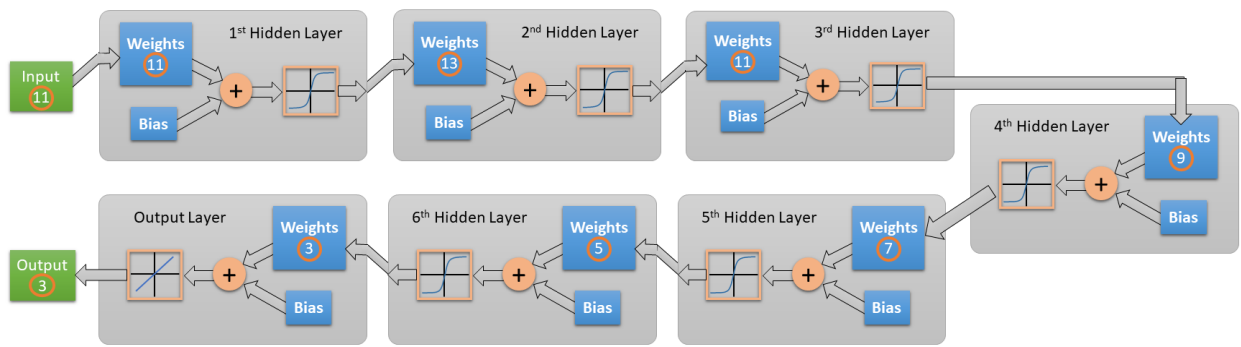

Figure 5: The used DL layers for the approximations of $a$, $b$, and $c$ with TOs.

To apply a DL model, we randomly divided the data into three sets: training, validation, and testing, according to the proportions of 0.85, 0.135, and 0.015, respectively. To achieve optimal results from our DL model, we first normalized the training data and stored the mean and variance of the training samples. We need these values to revert the output back to its original scale after completing the DL process.

| Case | Name | Type | Kurtosis | $\zeta$ | $\vartheta$ | $\iota$ | $\xi$ |
|---|---|---|---|---|---|---|---|
| 1 | Uniform | Bounded | $-1.2$ | 0 | 0.6465 | $-1.8153$ | 3.6306 |
| 2 | Triangular | Bounded | $-0.6$ | 0 | 1.3983 | $-3.1097$ | 6.2195 |
| 3 | Normal | Unbounded | 0 | 0 | 100 | 0 | 100 |
| 4 | Student T-10 | Unbounded | $+1.0$ | 0 | 2.3212 | 0 | 2.1094 |
| 5 | Student T-6 | Unbounded | $+3.0$ | 0 | 1.6104 | 0 | 1.3118 |
| 6 | Student T-5 | Unbounded | $+6.0$ | 0 | 1.3493 | 0 | 1 |

Table 1: Johnson-type distributions benchmark related to Equations (1.1) and (1.2).

The process of selecting a suitable DL model can be somehow challenging. We outline the steps for choosing an appropriate model in Figure 4. To enhance the DL model, we can use methods such as Bayesian optimization for hyperparameter tuning, regularization, and other available techniques. The optimal model we identified for the existing data is presented in Figure 5. In this model, we have 11 input neurons along with 6 hidden layers. Each hidden layer contains a different number of neurons, but all utilize the hyperbolic tangent activation function. Since our problem is a regression task, we choose a regression activation function for the output layer, which consists of three neurons corresponding to $a$, $b$, and $c$.

The next important step involves optimizing the weights and biases of the model. To achieve this, we employed the stochastic Gradient Descend (SGD) optimizer with 10,000 epochs, an initial learning rate of 0.02, a momentum of 0.9, and a validation frequency of 10, along with the Mean Squared Error (MSE) loss function.

We calculate the Root Mean Squared Error (RMSE) for the training and validation sets to evaluate the model. Then, we calculate the Mean Absolute Error (MAE) for the validation set and the $R^2$ values for the regressions of the test set, separately for $a$, $b$, and $c$. To assess overfitting, we also calculate the Mean Squared Error (MSE) for all three sets designated with indices 1 to 3. Here are the corresponding values:

$$\text{RMSE} = 0.0761, \begin{cases} \text{MAE}_a = 0.0183 \\ \text{MAE}_b = 0.0280, \\ \text{MAE}_c = 0.0136 \end{cases} \begin{cases} R_a^2 = 0.9997 \\ R_b^2 = 0.9965, \\ R_c^2 = 0.9963 \end{cases} \begin{cases} MSE_1 = 0.2604 \\ MSE_2 = 0.0019. \\ MSE_3 = 0.0018 \end{cases}$$

Based on these values, we observe that the model performs satisfactorily on the data. Additionally, there is no indication of overfitting, as MSE values are ordered such that $MSE_1 > MSE_2 > MSE_3$ for the training, validation, and test sets, respectively.

We draw the regression lines for the training, validation, and test sets in Figure 6. Almost all predicted points coincide perfectly with their corresponding actual values. We perform the inverse normalization process on the output data for the test set. Then, using this modified data, we plot the predicted density function of SND along with the simulated and the manually calculated SND densities in Figure 7 to compare the accuracy of the results. The summed density of the estimated $f_{SR_{t,N}}(\cdot)$ slightly deviates from 1. To ensure that the total probabilities sum to 1, we normalize the values by dividing them by the total sum. This adjustment confirms the properties of density functions.

An important final point to note about the DL model is that we constructed this model for two sample sizes $n = 20,50$ and TOs that are OOC. Consequently, this model has learned how to estimate the three SND parameters under these conditions. If we wish to expand the model, we need to incorporate the relevant values associated with these scenarios into the main dataset. However, since we typically utilize an ordinary ND for all UOs and TOs under IC conditions, there is no need to add this data for the DL process.

For computations of $ARL$s, we need control limits $(-C, C)$ for UOs and TOs, separately. Although, in both configurations the process is IC and we use ND, the mean and variance formulas are different for UOs and TOs. The desired confidence level is %99.73. The value that satisfies (2.6) either with $m_1$ and $\mu_2$ for UOs or $m'_1$ and $\mu'_2$ for TOs is the desired $C$. Since we have two $n$ we have to calculate $C$ two times for UOs and TOs with $p = 0.5$. The reason for choosing $p = 0.5$ is that, in the absence of ties and shifts in a process, considering all distributions in Table 1 and Equation (3.3), we have $p_0 = 0$ and $p_{-1} = p_{+1} = 0.5$. $C$s for $n = 20, 50$ are:

$$\text{UO:} \begin{cases} C = 162, n = 20 \\ C = 623, n = 50 \end{cases} \quad \text{TO:} \begin{cases} C = 64, n = 20 \\ C = 231, n = 50 \end{cases}.$$

We calculate the $ARL$s for $n = 20$ and $n = 50$ in Tables 3 and 4, respectively. $ARL_0$ values are placed in the center columns of the tables, while the sides of the columns contain the $ARL_1$ values for various shifts. The $ARL_1$ values exhibit near symmetry for both positive and negative shifts. The size of $ARL_0$ represents the number of samples required for a sample to be erroneously detected as OOC. Conversely, $ARL_1$ indicates the number of samples needed for the system to correctly identify changes in the production process and measurements. Therefore, as we previously mentioned, a larger $ARL_0$ results in delayed false alarms, while a smaller $ARL_1$ allows the process to more quickly detect unwanted changes in production. Both of these strategies help prevent the wastage of time and resources.

$ARL_0$s are sufficiently large for UOs and TOs in Tables 3 and 4. For TOs, the values of $ARL_1$ are small when considering $\delta = 0.1$ and $\delta = -0.1$ in both tables for $n = 20$ and $n = 50$. In contrast, the situation for UOs is different; the values of $ARL_1$ begin to decrease slightly for small $\delta$. Additionally, for the largest $\delta$ values mentioned in the tables, the $ARL_1$ values become small. Therefore, when ties are present, the process detects changes more quickly and is sensitive to small variations. In contrast, when no tie exists, the process identifies undesirable changes at a slower pace than in the tie-present scenario. This difference in sensitivity and responsiveness can significantly impact the efficiency of monitoring and controlling the production process.

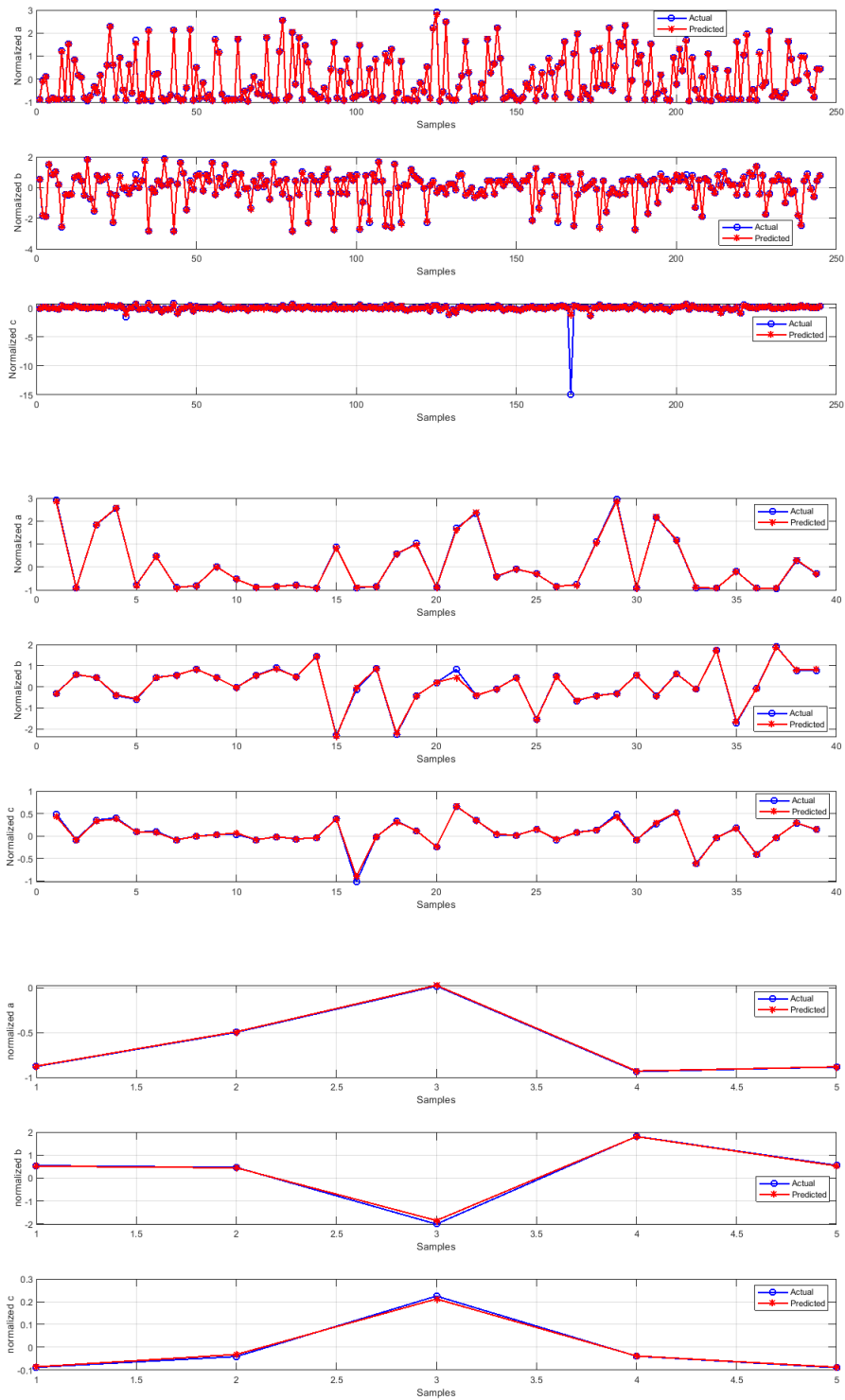

*Figure 6: The regression plots of the actual and predicted values for the training, validation, and test sets are displayed from top to bottom.*

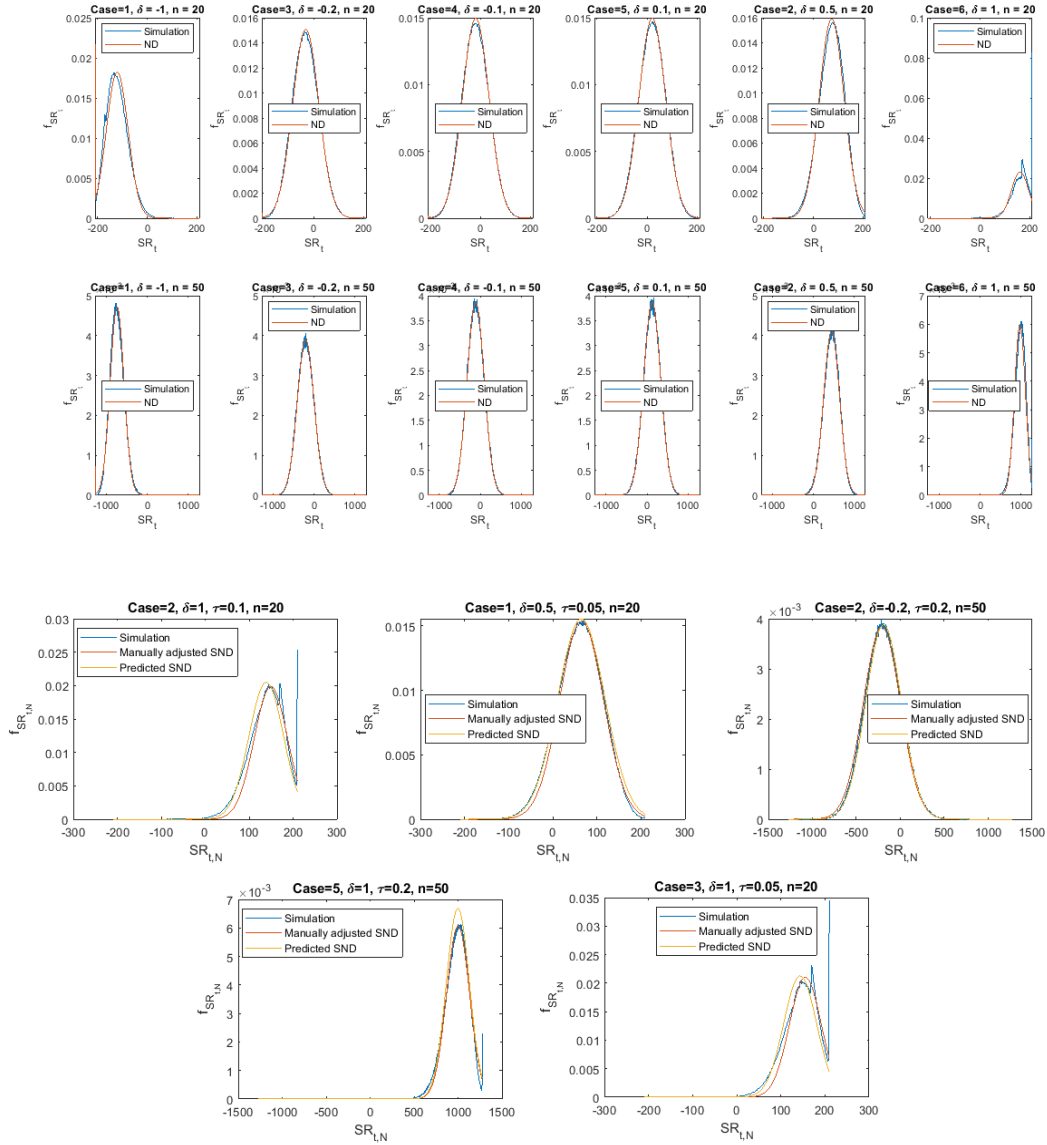

*Figure 7: The top twelve plots display the simulated and approximated densities of UOs, with specific settings detailed for each plot. The remaining five plots illustrate the TOs, showcasing the simulated densities, manually adjusted SND, and the DL predictions of SND.*

| $\tau$ | Case | Prob | | | | | $\delta$ | | | | |
|---|---|---|---|---|---|---|---|---|---|---|---|
| | | | $-1$ | $-0.5$ | $-0.2$ | $-0.1$ | $0$ | $0.1$ | $0.2$ | $0.5$ | $1$ |
| 0 | 1 | $p_{-1}$ | 0.7885 | 0.6427 | 0.5569 | 0.5284 | 0.5 | 0.4716 | 0.4431 | 0.3573 | 0.2115 |
| | | $p_{+1}$ | 0.2115 | 0.3573 | 0.4431 | 0.4716 | 0.5 | 0.5284 | 0.5569 | 0.6427 | 0.7885 |
| | 2 | $p_{-1}$ | 0.8244 | 0.6749 | 0.5714 | 0.5358 | 0.5 | 0.4641 | 0.4285 | 0.325 | 0.1756 |
| | | $p_{+1}$ | 0.1756 | 0.3251 | 0.4286 | 0.4642 | 0.5 | 0.5359 | 0.5715 | 0.675 | 0.8244 |

| τ | Case | Prob | | | | | δ | | | | |
|---|---|---|---|---|---|---|---|---|---|---|---|
| | 3 | $p_{-1}$ | 0.8413 | 0.6915 | 0.5793 | 0.5398 | 0.5 | 0.4602 | 0.4207 | 0.3085 | 0.1587 |
| | | $p_{+1}$ | 0.1587 | 0.3085 | 0.4207 | 0.4602 | 0.5 | 0.5398 | 0.5793 | 0.6915 | 0.8413 |
| | 4 | $p_{-1}$ | 0.8561 | 0.7072 | 0.587 | 0.5438 | 0.5 | 0.4562 | 0.413 | 0.2928 | 0.1439 |
| | | $p_{+1}$ | 0.1439 | 0.2928 | 0.413 | 0.4562 | 0.5 | 0.5438 | 0.587 | 0.7072 | 0.8561 |
| | 5 | $p_{-1}$ | 0.8712 | 0.7257 | 0.5966 | 0.5488 | 0.5 | 0.4512 | 0.4034 | 0.2743 | 0.1288 |
| | | $p_{+1}$ | 0.1288 | 0.2743 | 0.4034 | 0.4512 | 0.5 | 0.5488 | 0.5966 | 0.7257 | 0.8712 |
| | 6 | $p_{-1}$ | 0.8828 | 0.7419 | 0.6057 | 0.5536 | 0.5 | 0.4464 | 0.3943 | 0.2581 | 0.1172 |
| | | $p_{+1}$ | 0.1172 | 0.2581 | 0.3943 | 0.4464 | 0.5 | 0.5536 | 0.6057 | 0.7419 | 0.8828 |
| 0.05 | 1 | $p_{-1}$ | 0.7811 | 0.6355 | 0.5497 | 0.5213 | 0.4929 | 0.4645 | 0.436 | 0.3502 | 0.2041 |
| | | $p_{+1}$ | 0.2041 | 0.3502 | 0.436 | 0.4645 | 0.4929 | 0.5213 | 0.5497 | 0.6355 | 0.7811 |
| | 2 | $p_{-1}$ | 0.8179 | 0.6666 | 0.5626 | 0.5269 | 0.491 | 0.4552 | 0.4197 | 0.3168 | 0.1691 |
| | | $p_{+1}$ | 0.1692 | 0.3168 | 0.4197 | 0.4552 | 0.491 | 0.5269 | 0.5626 | 0.6666 | 0.8179 |
| | 3 | $p_{-1}$ | 0.8352 | 0.6826 | 0.5695 | 0.5299 | 0.49 | 0.4503 | 0.411 | 0.2998 | 0.1527 |
| | | $p_{+1}$ | 0.1527 | 0.2998 | 0.411 | 0.4503 | 0.49 | 0.5299 | 0.5695 | 0.6826 | 0.8352 |
| | 4 | $p_{-1}$ | 0.8504 | 0.6979 | 0.5763 | 0.5329 | 0.489 | 0.4453 | 0.4024 | 0.2837 | 0.1384 |
| | | $p_{+1}$ | 0.1384 | 0.2837 | 0.4024 | 0.4453 | 0.489 | 0.5329 | 0.5763 | 0.6979 | 0.8504 |
| | 5 | $p_{-1}$ | 0.866 | 0.716 | 0.5848 | 0.5367 | 0.4878 | 0.4391 | 0.3917 | 0.2649 | 0.1238 |
| | | $p_{+1}$ | 0.1238 | 0.2649 | 0.3917 | 0.4391 | 0.4878 | 0.5367 | 0.5848 | 0.716 | 0.866 |
| | 6 | $p_{-1}$ | 0.878 | 0.732 | 0.5929 | 0.5403 | 0.4865 | 0.4332 | 0.3817 | 0.2485 | 0.1126 |
| | | $p_{+1}$ | 0.1126 | 0.2485 | 0.3817 | 0.4332 | 0.4865 | 0.5403 | 0.5929 | 0.732 | 0.878 |
| 0.1 | 1 | $p_{-1}$ | 0.7737 | 0.6283 | 0.5426 | 0.5426 | 0.4858 | 0.4858 | 0.4289 | 0.343 | 0.1967 |
| | | $p_{+1}$ | 0.1967 | 0.343 | 0.4289 | 0.4289 | 0.4858 | 0.4858 | 0.5426 | 0.6283 | 0.7737 |
| | 2 | $p_{-1}$ | 0.8112 | 0.6582 | 0.5537 | 0.5537 | 0.482 | 0.482 | 0.4109 | 0.3086 | 0.1628 |
| | | $p_{+1}$ | 0.1629 | 0.3086 | 0.4109 | 0.4109 | 0.4821 | 0.4821 | 0.5537 | 0.6582 | 0.8113 |
| | 3 | $p_{-1}$ | 0.8289 | 0.6736 | 0.5596 | 0.5596 | 0.4801 | 0.4801 | 0.4013 | 0.2912 | 0.1469 |
| | | $p_{+1}$ | 0.1469 | 0.2912 | 0.4013 | 0.4013 | 0.4801 | 0.4801 | 0.5596 | 0.6736 | 0.8289 |
| | 4 | $p_{-1}$ | 0.8445 | 0.6885 | 0.5655 | 0.5655 | 0.4781 | 0.4781 | 0.3919 | 0.2747 | 0.133 |
| | | $p_{+1}$ | 0.133 | 0.2747 | 0.3919 | 0.3919 | 0.4781 | 0.4781 | 0.5655 | 0.6885 | 0.8445 |
| | 5 | $p_{-1}$ | 0.8606 | 0.7061 | 0.5729 | 0.5729 | 0.4755 | 0.4755 | 0.3802 | 0.2556 | 0.1189 |
| | | $p_{+1}$ | 0.1189 | 0.2556 | 0.3802 | 0.3802 | 0.4755 | 0.4755 | 0.5729 | 0.7061 | 0.8606 |
| | 6 | $p_{-1}$ | 0.8731 | 0.7219 | 0.5799 | 0.5799 | 0.4731 | 0.4731 | 0.3692 | 0.2392 | 0.1082 |
| | | $p_{+1}$ | 0.1082 | 0.2392 | 0.3692 | 0.3692 | 0.4731 | 0.4731 | 0.5799 | 0.7219 | 0.8731 |
| 0.2 | 1 | $p_{-1}$ | 0.759 | 0.614 | 0.5284 | 0.5 | 0.4716 | 0.4431 | 0.4146 | 0.3285 | 0.1819 |
| | | $p_{+1}$ | 0.1819 | 0.3285 | 0.4146 | 0.4431 | 0.4716 | 0.5 | 0.5284 | 0.614 | 0.759 |
| | 2 | $p_{-1}$ | 0.7976 | 0.6412 | 0.5358 | 0.5 | 0.4641 | 0.4285 | 0.3933 | 0.2924 | 0.1506 |

| $\tau$ | Case | Prob | | | | $\delta$ | | | | | |
|---|---|---|---|---|---|---|---|---|---|---|---|
| 3 | | $p_{+1}$ | 0.1506 | 0.2924 | 0.3934 | 0.4286 | 0.4642 | 0.5 | 0.5359 | 0.6412 | 0.7976 |
| | | $p_{-1}$ | 0.8159 | 0.6554 | 0.5398 | 0.5 | 0.4602 | 0.4207 | 0.3821 | 0.2743 | 0.1357 |
| 4 | | $p_{+1}$ | 0.1357 | 0.2743 | 0.3821 | 0.4207 | 0.4602 | 0.5 | 0.5398 | 0.6554 | 0.8159 |
| | | $p_{-1}$ | 0.8321 | 0.6691 | 0.5438 | 0.5 | 0.4562 | 0.413 | 0.3711 | 0.2573 | 0.1227 |
| 5 | | $p_{+1}$ | 0.1227 | 0.2573 | 0.3711 | 0.413 | 0.4562 | 0.5 | 0.5438 | 0.6691 | 0.8321 |
| | | $p_{-1}$ | 0.8491 | 0.6857 | 0.5488 | 0.5 | 0.4512 | 0.4034 | 0.3575 | 0.2379 | 0.1097 |
| 6 | | $p_{+1}$ | 0.1097 | 0.2379 | 0.3575 | 0.4034 | 0.4512 | 0.5 | 0.5488 | 0.6857 | 0.8491 |
| | | $p_{-1}$ | 0.8625 | 0.7007 | 0.5536 | 0.5 | 0.4464 | 0.3943 | 0.345 | 0.2214 | 0.0999 |
| | | $p_{+1}$ | 0.0999 | 0.2214 | 0.345 | 0.3943 | 0.4464 | 0.5 | 0.5536 | 0.7007 | 0.8625 |

*Table 2: The probabilities of occurrence for $+1$ and $-1$, based on the densities presented in Table 1 for UO and TO. The sum of $p_{-1}$ and $p_{+1}$ equals 1 for UO, while for TO, it is less than 1.*

# Conclusion

Statistical quality control methods are widely employed in many manufacturing processes to minimize financial losses. CCs are the most effective tools for promptly monitoring and identifying correct shifts. possible. The presumption of the SCC pertains to the distribution of data obtained from a production process. The distribution must be exactly or approximately normal; however, this condition is not valid in many processes. Furthermore, in certain processes, the measurement devices used to record the quantities of quality characteristics may lack sufficient accuracy. As a result, they can introduce rounding errors into the process dataset. These errors can lead to the creation of ties in the outputs. In such cases, the true values of observations are unrecorded, and instead, rounded numbers are recorded. Therefore, CC must to be that which effectively handle with ties in datasets.

In this paper, we redesign a SS-RCC for TOs and UOs. The challenge we face is that the distributions are unknown. To address this issue, we apply the Wilcoxon statistic. In the absence of a tie, we approximate the distribution of the statistic using the ND. When ties are present, we discretely approximate the distribution of the statistic using SND defined in this paper. This means that we estimate the probabilities for each point in the sample space within a small interval surrounding that point. In this case, we need to estimate the three newly defined parameters for SND. To accomplish this, we employ DL techniques. Then, we determine symmetric control limits based on an appropriate confidence level and the estimated distributions. Finally, we calculate $ARL$s for six symmetric Johnson-type distributions and perform a conditional simulation to encompass various scenarios. In conclusion, our proposed method demonstrates supreme effectiveness in detecting shifts in processes.

## Declarations

This work was supported by the Innovative Research Group Project of the National Natural Science Foundation of China (Grant No. 51975253). The authors report no conflicts of interest or personal ties that may affect the work in this paper. The contributions are as follows: Seyedeh Azadeh Fallah Mortezanejad: Methodology, Conceptualization, Validation, Writing-original draft, Writing-review and editing. Ruochen Wang: Supervision, Methodology, Investigation, Writing-original draft, Writing-review and editing.

# Calculations of $SR_{t,N}$

The probability generating function of the binomial random variable $N$ is given by:

$$P_N(t) = E(t^N) = \sum_{N=0}^{n} \binom{n}{N} t^N (1-p_0)^N p_0^{n-N} = (p_0 + (1-p_0)t)^n.$$

Some required expectations of $N$ are:

$$\begin{cases} E(N) = & n(1-p_0) \\ E(N^2) = & n(n-1)(1-p_0)^2 + E(N) \\ E(N^3) = & n(n-1)(n-2)(1-p_0)^3 + 3E(N^2) - 2E(N) \\ E(N^4) = & n(n-1)(n-2)(n-3)(1-p_0)^4 + 6E(N^3) - 11E(N^2) + 6E(N) \\ E(N^5) = & n(n-1)(n-2)(n-3)(n-4)(1-p_0)^5 + 10E(N^4) - 35E(N^3) \\ & +50E(N^2) - 24E(N) \\ E(N^6) = & n(n-1)(n-2)(n-3)(n-4)(n-5)(1-p_0)^6 \\ & +15E(N^5) - 85E(N^4) + 225E(N^3) - 274E(N^2) + 120E(N) \\ E(N^7) = & n(n-1)(n-2)(n-3)(n-4)(n-5)(n-6)(1-p_0)^7 \\ & +21E(N^6) - 175E(N^5) + 735E(N^4) - 1624E(N^3) \\ & +1764E(N^2) - 720E(N) \\ E(N^8) = & n(n-1)(n-2)(n-3)(n-4)(n-5)(n-6)(n-7)(1-p_0)^8 \\ & +28E(N^7) - 322E(N^6) + 1960E(N^5) - 6769E(N^4) \\ & +13132E(N^3) - 13068E(N^2) + 5040E(N) \end{cases}.$$

These are some expectations of $SR_{t,N}$ for $k, j, i, l = 1, \cdots, n$ that we need to address below:

$$\begin{cases} E(SR_{t,N}^3) & = m_1(S'_{t,k}) = \pi_{+1} - \pi_{-1} \\ E(SR_{k_{t,N}}^2 SR_{j_{t,N}}) & = \pi_{+1} - \pi_{-1} \quad \text{for } k \neq j \\ E(SR_{t,N}^4) & = E(S_{t,k}^{\prime 2}) = 1 \\ E(SR_{k_{t,N}}^2 SR_{j_{t,N}}^2) & = 1 \quad \text{for } k \neq j \\ E(SR_{k_{t,N}} SR_{j_{t,N}}) & = (\pi_{+1} - \pi_{-1})^2 \quad \text{for } k \neq j \\ E(SR_{k_{t,N}}^3 SR_{j_{t,N}}) & = (\pi_{+1} - \pi_{-1})^2 \quad \text{for } k \neq j \\ E(SR_{k_{t,N}}^2 SR_{j_{t,N}} SR_{i_{t,N}}) & = (\pi_{+1} - \pi_{-1})^2 \quad \text{for } k \neq j \neq i \\ E(SR_{k_{t,N}} SR_{j_{t,N}} SR_{i_{t,N}}) & = (\pi_{+1} - \pi_{-1})^3 \quad \text{for } k \neq j \neq i \\ E(SR_{k_{t,N}} SR_{j_{t,N}} SR_{i_{t,N}} SR_{l_{t,N}}) & = (\pi_{+1} - \pi_{-1})^4 \quad \text{for } k \neq j \neq i \neq l \end{cases}.$$

The formulas to compute $SR_{t,N}^2$, $SR_{t,N}^3$ and $SR_{t,N}^4$ are:

$$SR_{t,N}^2 = \left(\sum_{k=1}^{N} k\, S'_{t,k}\right)^2 = \sum_{k=1}^{N} k^2\, S'^{2}_{t,k} + \sum_{\substack{k,j=1 \\ k \neq j}}^{N} k\, j\, S'_{t,k}\, S'_{t,j},$$

$$SR_{t,N}^3 = \sum_{k=1}^{N} k^3\, S'^{3}_{t,k} + 3 \sum_{\substack{k,j=1 \\ k \neq j}}^{N} k^2\, S'^{2}_{t,k}\, jS'_{t,j} + 6 \sum_{\substack{k,j,i=1 \\ i<j<k}}^{N} i\, S'_{t,i}\, jS'_{t,j}\, kS'_{t,k},$$

$$SR_{t,N}^4 = \sum_{k=1}^{N} k^4\, S'^{4}_{t,k} + 6 \sum_{\substack{k,j=1 \\ j<k}}^{N} k^2\, j^2 S'^{2}_{t,k} S'^{2}_{t,j} + 4 \sum_{\substack{k,j=1 \\ k \neq j}}^{N} k^3\, jS_{t,j} S'^{3}_{t,k}$$

$$+ 36 \sum_{\substack{k,j,i=1 \\ i \neq j \neq k}}^{N} i^2\, jk S'^{2}_{t,i} S'_{t,j} S'_{t,k} + 4 \sum_{\substack{k,j,i,l=1 \\ i<j<k<l}}^{N} i\, j\, k\, l\, S'_{t,i} S'_{t,j} S'_{t,k} S'_{t,l}.$$

To calculate $m'_2$ in detail, we need closed-forms of summations for integers from 1 to the variable $N$ provided in 7. Then, we get:

$$m'_2 = m'_2(SR_{t,N}^2) = E\left(E\left[\left(\sum_{k=1}^{N} k\, S'_{t,k}\right)^2 \Big| N\right]\right)$$

$$= E\left(\sum_{k=1}^{N} k^2\, E(S'^2_{t,k}|N) + \sum_{\substack{k,j=1 \\ k \neq j}}^{N} k\, j\, E(S'_{t,k} S'_{t,j}|N)\right)$$

$$= \frac{1}{6} E(N(N+1)(2N+1)) + (\pi_{+1} - \pi_{-1})^2 E\left(\frac{N^2(N+1)^2}{4} - \frac{N(N+1)(2N+1)}{6}\right)$$

$$= E\left(\frac{1}{3}N^3 + \frac{1}{2}N^2 + \frac{1}{6}N\right) + (\pi_{+1} - \pi_{-1})^2 E\left(\frac{1}{4}N^4 + \frac{1}{6}N^3 - \frac{1}{4}N^2 - \frac{1}{6}N\right).$$

Also, we have the same for $m'_3$ and $m'_4$:

$$m'_3 = m'_3(SR_{t,N}^3) = E\left(E\left[\left(\sum_{k=1}^{N} k\, S'_{t,k}\right)^3 \Big| N\right]\right)$$

$$= E\left(\sum_{k=1}^{N} k^3\, E(S'^3_{t,k}|N) + 3\sum_{\substack{k,j=1 \\ k \neq j}}^{N} k^2\, j\, E(S'^2_{t,k} S'_{t,j}|N) + 6\sum_{\substack{k,j,i=1 \\ i<j<k}}^{N} i\, j\, k\, E(S'_{t,i} S'_{t,j} S'_{t,k}|N)\right)$$

$$= (\pi_{+1} - \pi_{-1}) E\left(\sum_{k=1}^{N} k^3\right) + 3(\pi_{+1} - \pi_{-1}) E\left(\sum_{\substack{k,j=1 \\ k \neq j}}^{N} k^2\, j\right) + 6(\pi_{+1} - \pi_{-1})^3 E\left(\sum_{\substack{k,j,i=1 \\ i<j<k}}^{N} i\, j\, k\right)$$

$$= (\pi_{+1} - \pi_{-1}) E\left(\frac{N^2(N+1)^2}{4}\right) + 3(\pi_{+1} - \pi_{-1}) E\left(\frac{N^2(N+1)^2(N-1)}{6}\right)$$

$$+ 6(\pi_{+1} - \pi_{-1})^3 E\left(\frac{N^2(N+1)^2(N^2 - 3N + 2)}{48}\right)$$

$$= \frac{\pi_{+1} - \pi_{-1}}{4} E(2N^5 + 3N^4 - N^2) + \frac{(\pi_{+1} - \pi_{-1})^3}{8} E(N^6 - N^5 - 3N^4 + N^3 + 2N^2),$$

$$m'_4 = m'_4(SR^4_{t,N}) = E\left(E\left[\left(\sum_{k=1}^N k\, S'_{t,k}\right)^4 \Big| N\right]\right)$$

$$= E\left(\sum_{k=1}^N k^4\, E(S'^4_{t,k}|N) + 6 \sum_{\substack{k,j=1 \\ j<k}}^N k^2\, j^2 E(S'^2_{t,k} S'^2_{t,j}|N) + 4 \sum_{\substack{k,j=1 \\ k\neq j}}^N k^3\, j E(S'_{t,j} S'^3_{t,k}|N)\right.$$

$$\left. +36 \sum_{\substack{k,j,i=1 \\ i\neq j\neq k}}^N i^2\, jk E(S'^2_{t,i} S'_{t,j} S'_{t,k}|N) + 4 \sum_{\substack{k,j,i,l=1 \\ i<j<k<l}}^N i\, jkl E(S'_{t,i} S'_{t,j} S'_{t,k} S'_{t,l}|N)\right)$$

$$= E\left(\frac{1}{3}N^6 + \frac{3}{5}N^5 + \frac{1}{12}N^4 - \frac{1}{6}N^3 + \frac{1}{12}N^2 + \frac{1}{15}N\right)$$

$$+ \frac{(\pi_{+1} - \pi_{-1})^3}{30} E(15N^6 + 21N^5 - 15N^4 - 25N^3 + 4N)$$

$$+ \frac{(\pi_{+1} - \pi_{-1})^2}{10} E(30N^7 - 25N^6 - 111N^5 + 35N^4 + 105N^3 - 10N^2 - 24N)$$

$$+ \frac{(\pi_{+1} - \pi_{-1})^4}{1440} E(15N^8 - 60N^7 - 10N^6 + 192N^5 - 25N^4 - 180N^3 + 20N^2 + 48N).$$

We calculate $\mu'_2$, $\mu'_3$, and $\mu'_4$ in the following:

$$\mu'_2 = m'_2 - m'^2_1$$

$$= \frac{1}{2} E(N^2 + N) + (\pi_{+1} - \pi_{-1})^2 E\left(\frac{1}{4}N^4 + \frac{1}{6}N^3 - \frac{1}{4}N^2 - \frac{1}{6}N\right) - \frac{(\pi_{+1} - \pi_{-1})^2}{4} E(N^2 + N)^2,$$

$$\mu'_3 = m'_3 - 3m'_1 m'_2 + 2m'^3_1$$

$$= \frac{\pi_{+1} - \pi_{-1}}{4} E(2N^5 + 3N^4 - N^2) - \frac{3(\pi_{+1} - \pi_{-1})}{4} E(N^2 + N)^2$$

$$+ \frac{(\pi_{+1} - \pi_{-1})^3}{8} E(N^6 - N^5 - 3N^4 + N^3 + 2N^2)$$

$$- \frac{3(\pi_{+1} - \pi_{-1})^3}{2} E\left(\frac{1}{4}N^4 + \frac{1}{6}N^3 - \frac{1}{4}N^2 - \frac{1}{6}N\right) + \frac{(\pi_{+1} - \pi_{-1})^3}{4} E(N^2 + N)^3$$

$$= \frac{\pi_{+1} - \pi_{-1}}{4} \left(E(2N^5 + 3N^4 - N^2) - 3E(N^2 + N)^2\right)$$

$$+ \frac{(\pi_{+1} - \pi_{-1})^3}{8} \left(E(N^6 - N^5 - 6N^4 - N^3 + 5N^2 + 2N) + 2E(N^2 + N)^3\right),$$

$$\begin{aligned}
\mu'_4 &= -3m'^4_1 + 6m'^2_1 m'_2 - 4m'_1 m'_3 + m'_4 \\
&= -\frac{3(\pi_{+1} - \pi_{-1})^4}{16} E(N^2 + N)^4 + \frac{3(\pi_{+1} - \pi_{-1})^2}{4} E(N^2 + N)^3 \\
&+ \frac{3(\pi_{+1} - \pi_{-1})^4}{2} E(N^2 + N)^2 E\left(\frac{1}{4}N^4 + \frac{1}{6}N^3 - \frac{1}{4}N^2 - \frac{1}{6}N\right) \\
&- \frac{(\pi_{+1} - \pi_{-1})^2}{2} E(N^2 + N) E(2N^5 + 3N^4 - N^2) \\
&- \frac{(\pi_{+1} - \pi_{-1})^4}{4} E(N^2 + N) E(N^6 - N^5 - 3N^4 + N^3 + 2N^2) \\
&+ E\left(\frac{1}{3}N^6 + \frac{3}{5}N^5 + \frac{1}{12}N^4 - \frac{1}{6}N^3 + \frac{1}{12}N^2 + \frac{1}{15}N\right) \\
&+ \frac{(\pi_{+1} - \pi_{-1})^3}{30} E(15N^6 + 21N^5 - 15N^4 - 25N^3 + 4N) \\
&+ \frac{(\pi_{+1} - \pi_{-1})^2}{10} E(30N^7 - 25N^6 - 111N^5 + 35N^4 + 105N^3 - 10N^2 - 24N) \\
&+ \frac{(\pi_{+1} - \pi_{-1})^4}{1440} E(15N^8 - 60N^7 - 10N^6 + 192N^5 - 25N^4 - 180N^3 + 20N^2 + 48N) \\[4pt]
&= \frac{(\pi_{+1} - \pi_{-1})^4}{1440} \{2160 E(N^2 + N)^2 E\left(\frac{1}{4}N^4 + \frac{1}{6}N^3 - \frac{1}{4}N^2 - \frac{1}{6}N\right) \\
&\quad - 270 E(N^2 + N)^4 - 360 E(N^2 + N) E(N^6 - N^5 - 3N^4 + N^3 + 2N^2) \\
&\quad + E(15N^8 - 60N^7 - 10N^6 + 192N^5 - 25N^4 - 180N^3 + 20N^2 + 48N)\} \\
&+ \frac{(\pi_{+1} - \pi_{-1})^3}{30} E(15N^6 + 21N^5 - 15N^4 - 25N^3 + 4N) \\
&+ \frac{(\pi_{+1} - \pi_{-1})^2}{10} \{\frac{15}{2} E(N^2 + N)^3 - 5 E(N^2 + N) E(2N^5 + 3N^4 - N^2) \\
&\quad + E(30N^7 - 25N^6 - 111N^5 + 35N^4 + 105N^3 - 10N^2 - 24N)\} \\
&+ E\left(\frac{1}{3}N^6 + \frac{3}{5}N^5 + \frac{1}{12}N^4 - \frac{1}{6}N^3 + \frac{1}{12}N^2 + \frac{1}{15}N\right).
\end{aligned}$$

## Appendix B: Some Series Computations

In this appendix, we aim to present several summations along with their closed-forms. We require these to compute the moments of $SR_{t,n}$. We express the summations of the $s$th powers of the first $n$ integers as follows:

$$\sum_{k=1}^{n} k = \frac{n(n+1)}{2},$$

$$\sum_{k=1}^{n} k^2 = \frac{n(n+1)(2n+1)}{6},$$

$$\sum_{k=1}^{n} k^3 = \frac{n^2(n+1)^2}{4},$$

$$\sum_{k=1}^{n} k^4 = \frac{n^5}{5} + \frac{n^4}{2} + \frac{n^3}{3} - \frac{n}{30}.$$

We formulate some summations binary products of integers from 1 to $n$ as follows:

$$\sum_{\substack{k,j=1 \\ k \neq j}}^{n} k\,j = \left(\frac{n(n+1)}{2}\right)^2 - \frac{n(n+1)(2n+1)}{6},$$

$$\sum_{\substack{k,j=1 \\ k \neq j}}^{n} k^2\,j = \frac{n^2(n+1)^2(n-1)}{6},$$

$$\sum_{\substack{k,j=1 \\ j<k}}^{n} k^2\,j^2 = \frac{n}{360}(20n^5 + 24n^4 - 25n^3 - 30n^2 + 5n + 6),$$

$$\sum_{\substack{k,j=1 \\ k \neq j}}^{n} k^3\,j = \frac{n}{120}(15n^5 + 21n^4 - 15n^3 - 25n^2 + 4),$$

where we have:

$$\sum_{\substack{k,j=1 \\ k \neq j}}^{n} k\,j = 2 \sum_{\substack{k,j=1 \\ k<j}}^{n} k\,j.$$

Here are some triple products of the integers:

$$\sum_{\substack{k,j,i=1 \\ i<j<k}}^{n} i\,j\,k = \frac{1}{48} n^2(n+1)^2(n^2 - 3n + 2),$$

$$\sum_{\substack{k,j,i=1 \\ i \neq j \neq k}}^{n} i^2\,j\,k = \frac{n}{360}(30n^6 - 25n^5 - 111n^4 + 35n^3 + 105n^2 - 10n - 24).$$

The quadruple product when $i < j < k < l$ is:

$$\sum_{\substack{k,j,i,l=1 \\ i<j<k<l}}^{n} ijkl = \frac{n}{5760}(15n^7 - 60n^6 - 10n^5 + 192n^4 - 25n^3 - 180n^2 + 20n + 48).$$